\documentclass[conference]{IEEEtran}

\usepackage[table,xcdraw]{xcolor}
\usepackage[english]{babel}
\usepackage{float}
\usepackage{color}
\usepackage{bbm}
\usepackage{amsmath, bm}
\usepackage{multirow}
\usepackage{epsfig}
\usepackage{amsmath}
\usepackage{mathtools}
\usepackage{amssymb}
\usepackage{verbatim}
\usepackage{lipsum}
\usepackage{graphicx}
\usepackage{cuted}
\usepackage{cite}
\usepackage{ulem}
\usepackage{url}
\usepackage{amsmath}
\usepackage{algorithm}
\usepackage[noend]{algpseudocode}
\usepackage{caption}
\usepackage{subfig}
\usepackage{mathtools}
\usepackage{authblk}
\DeclarePairedDelimiter{\ceil}{\lceil}{\rceil}

\graphicspath{ {./FIGS/} }

\newcommand{\ianNotes}{\textcolor{black}}

\title{A Competitive Edge: Can FPGAs Beat GPUs at DCNN Inference Acceleration in Resource-Limited Edge Computing Applications?}

\author[1]{Ian Colbert}
\author[1]{Jake Daly}
\author[1]{Ken Kreutz-Delgado}
\author[2]{Srinjoy Das}
\affil[1]{\footnotesize{Department of Electrical and Computer Engineering, University of California, San Diego}}
\affil[2]{\footnotesize{Department of Mathematics, University of California, San Diego}}

\begin{document}

\maketitle

\vspace*{-1cm}
\begin{abstract}
When trained as generative models, Deep Learning algorithms have shown exceptional performance on tasks involving high dimensional data such as image denoising and super-resolution.
In an increasingly connected world dominated by mobile and edge devices, there is surging demand for these algorithms to run locally on embedded platforms.
FPGAs, by virtue of their reprogrammability and low-power characteristics, are ideal candidates for these edge computing applications.
As such, we design a spatio-temporally parallelized hardware architecture capable of accelerating a deconvolution algorithm optimized for power-efficient inference on a resource-limited FPGA.
We propose this FPGA-based accelerator to be used for Deconvolutional Neural Network (DCNN) inference in low-power edge computing applications.
To this end, we develop methods that systematically exploit \ianNotes{micro-architectural innovations, design space exploration, and statistical analysis}.
Using a Xilinx PYNQ-Z2 FPGA, we leverage our architecture to accelerate inference for two DCNNs trained on the MNIST and CelebA datasets using the Wasserstein GAN framework.
On these networks, our FPGA design achieves a higher throughput to power ratio with lower run-to-run variation when compared to the NVIDIA Jetson TX1 edge computing GPU. 
\end{abstract}

\section{Introduction}

Generative models are widely used as a means of parameterizing distributions of high-dimensional signals and structures.
Among the various types of generative models, the Generative Adversarial Network (GAN) first proposed by Goodfellow \textit{et al.}~\cite{goodfellow2014generative} yields superior performance on applications such as image generation, super resolution, and language modeling~\cite{pan2019recent}.
The learning strategy of the GAN jointly optimizes a generator $G$ and a discriminator $D$. 
While the generator $G$ is trained to minimize the distance between the ground truth distribution $P_g$ and the model-parameterized distribution $P_\theta$, the discriminator $D$ is trained to separate $P_g$ from $P_\theta$.
Although training optimizes both $G$ and $D$, only the generator $G$ is needed for inference when drawing samples from $P_\theta$.

The typical GAN framework shown in Fig.~\ref{fig:gan} involves convolution layers, where $D$ is a Convolutional Neural Network (CNN) and $G$ is a Deconvolutional Neural Network (DCNN).
Traditionally, these networks are deployed on CPUs and GPUs using cloud computing infrastructures.
However, the proliferation of applications for mobile and edge computing have created new opportunities to deploy these models on embedded hardware for local inference.
\ianNotes{In contrast to CPUs and GPUs, 
FPGAs offer large-scale fine-grained parallelism and provide consistent power-efficient throughput, making them well-suited for these edge computing applications~\cite{biookaghazadeh2018fpgas}.}

\ianNotes{In this paper, we consider DCNN inference acceleration using a resource-limited Xilinx PYNQ-Z2 FPGA.
We benchmark our implementation against the NVIDIA Jetson TX1 GPU, a processor heavily optimized for edge computing applications, and achieve a superior throughput to power ratio.}
The contributions of this paper are as follows:

\begin{itemize}
    \item Significant enhancements over the algorithm proposed by~\cite{zhang2017design} that reduce resource utilization, improve dataflow, and exploit memory hierarchy
    \item A spatio-temporally parallelized hardware architecture specifically designed to exploit these algorithmic innovations for power-efficient acceleration of DCNN inference
    \item An application of high-dimensional statistical analyses to balance the trade-off between hardware performance and generative quality when exploring network sparsity  
\end{itemize}

\begin{figure}[t]
    \centering
    \includegraphics[width=0.8\linewidth]{FIGS/GAN_IC.png}
    \caption{\small{\textbf{Generative Adversarial Network~\cite{goodfellow2014generative} Architecture.}\\After training on the cloud, we map generator $G$ onto local hardware for low-power inference at the edge.}}
    \label{fig:gan}
\end{figure}


\vspace{-0.15cm}

\section{Related Research}
\label{sec:related-work}

Previous works take architectural and algorithmic approaches to accelerating deconvolution workloads.
The authors in~\cite{yazdanbakhsh2018flexigan} and~\cite{yazdanbakhsh2018ganax} 
reformulate the deconvolution operation as a sparse convolution 
and build complex architectures that unify SIMD and MIMD execution models.
Wang \textit{et al.}~\cite{wang2019towards} also use the zero-insertion deconvolution algorithm, approaching the problem by parallelizing over a uniform 2D systolic array hardware architecture to accelerate both 2D and 3D DCNNs.
Liu \textit{et al.}~\cite{liu2018memory} propose a tiling method with a memory-efficient architecture that limits off-chip memory accesses at the cost of increased resource utilization via on-chip buffering.
Chang \textit{et al}.~\cite{chang2018energy,chang2020towards} propose an accelerator that transforms the deconvolution operation into a convolution (TDC), requiring $stride^2$ as many filters and potentially zero-padding the input and weight matrices.
To improve dataflow, Tu \textit{et al.}~\cite{tu2019accelerating} explore the on-chip re-stitching of the disjoint output feature maps resulting from the TDC method. Mao \textit{et al.}~\cite{mao2020fdna} adapt this method in a piecewise manner to handle the load-imbalance resulting from zero-padding at the cost of increased hardware complexity. 
The algorithm first proposed by Zhang \textit{et al.}~\cite{zhang2017design} avoids the zero-insertion and zero-padding requirements of the methods outlined above.
We adapt this algorithm to a parallel hardware architecture as described in Sections~\ref{sec:deconv-algorithm} and~\ref{sec:hardware-architecture}.



\begin{figure}[t]
\centering
\includegraphics[width=0.43\linewidth]{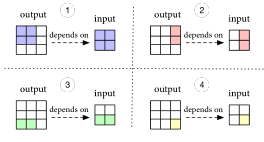}
\caption{\small{\textbf{Deconvolution Mapping of Input and Output Feature Maps.} Visualization from~\cite{zhang2017design} for mapping input and output blocks.}}
\label{fig:deconv-tetris}
\end{figure}

\begin{figure*}[h!]
\begin{center}
    \includegraphics[width=0.62\linewidth]{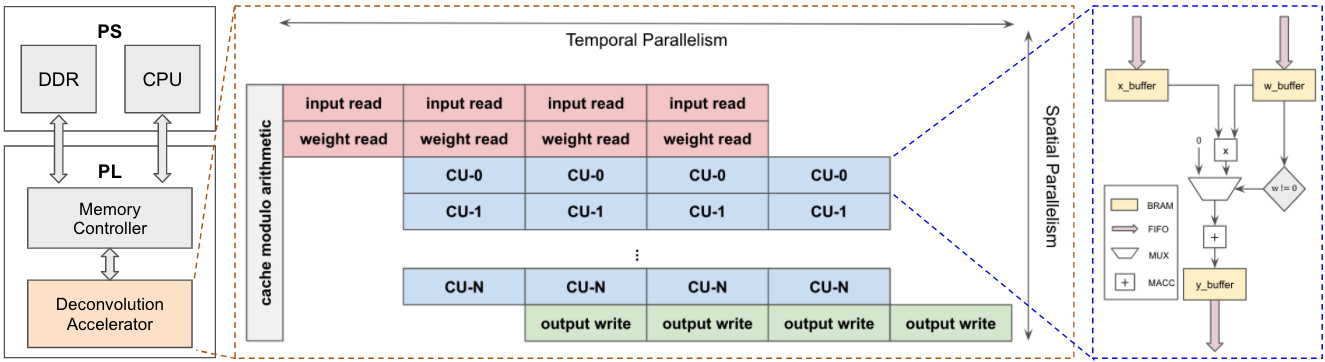}
\end{center}
\caption{ \protect\centering\small \protect\textbf{FPGA Hardware Architecture.} As discussed in Section~\protect\ref{sec:hardware-architecture}, we design a spatio-temporally parallelized hardware architecture customized to accelerate the deconvolution algorithm proposed in Section~\protect\ref{sec:deconv-algorithm} for low-power DCNN inference at the edge.}
\label{fig:hardware-architecture}
\end{figure*}

\section{Deconvolution Algorithm}
\label{sec:deconv-algorithm}


Standard deconvolution arithmetic traverses the input space, which requires a summation of regions that overlap in the output space~\cite{dumoulin2016guide}.
When realized in hardware, accumulating over these overlapping regions can require complex dataflow and increase resource utilization via on-chip buffering~\cite{zhang2017design,liu2018memory,chang2018energy}.
To circumvent this, Zhang \textit{et al.}~\cite{zhang2017design} redesign the deconvolution algorithm to directly loop over the output space at the cost of the expensive modulo arithmetic required to calculate dependent input pixels.
We propose the following three enhancements to adapt this reverse looping algorithm to a spatio-temporally parallelized hardware architecture. \\ 

\vspace{-0.25cm}
\noindent \textbf{(1) Preprocessing modulo arithmetic.} 
Standard deconvolution arithmetic calculates the indices of dependent output pixels $o_h$ from input index $i_h$ using weight index $k_h$, stride $S$, and padding $P$, as shown in Eq.~\ref{eq:output-pixel-calc}.
Here, tiling along the input space leads to overlapping blocks in the output space, creating communication overhead~\cite{zhang2017design,chang2018energy,tu2019accelerating}. 
\vspace{-0.1cm}
\begin{equation}
\small{o_h = i_h \times S + k_h - P}
\label{eq:output-pixel-calc}
\end{equation}
To avoid this, Zhang \textit{et al.}~\cite{zhang2017design} use the mapping in Fig.~\ref{fig:deconv-tetris} to loop over the output space and determine $i_h$ using Eq.~\ref{eq:input-pixel-calc}.
\begin{equation}
\small{i_h = \frac{o_h + P - k_h}{S}}
\label{eq:input-pixel-calc}
\end{equation}
When $S > 1$, Eq.~\ref{eq:input-pixel-calc} yields fractional values. To ensure functional correctness, Zhang \textit{et al.}~\cite{zhang2017design} propose a stride hole skipping technique, adding an offset value $f_h$ given by Eq.~\ref{eq:stride-hole-skipping}. 
\begin{equation}
\small{f_h = \textbf{mod}( S - \textbf{mod}( P - k_h, S ), S )}
\label{eq:stride-hole-skipping}
\end{equation}
However, the resulting input pixel calculation given by Eq.~\ref{eq:input-pixel-calc-final} relies on modulo arithmetic which increases resource utilization and power consumption when implemented in hardware.
\begin{equation}
\small{i_h = \frac{o_h + P - k_h + f_h}{S}}
\label{eq:input-pixel-calc-final}
\end{equation}
Observing that, in Eq.~\ref{eq:stride-hole-skipping}, $f_h$ is only dependent on $k_h$, we pre-compute and cache these offsets for each value of $k_h$.
This process reduces the number of modulo operations to $2K$, where $K$ is the weight filter size. This minimizes resource utilization and on-chip memory as $K$ tends to be small. \\

\begin{algorithm}[t]
\begin{algorithmic}
\small{
\State $\mathbf{y} \leftarrow \textbf{initializeToBias()}$
\For{$i_c = 0$, $i_c{+}{+}$, while $i_c < I_C$}
\State $\mathbf{x} \leftarrow \textbf{loadInputBlock()}$
\State $\mathbf{w} \leftarrow \textbf{loadWeightBlock()}$
\For{$k_h = 0$, $k_h{+}{+}$, while $k_h < K$}
\For{$k_w = 0$, $k_w{+}{+}$, while $k_w < K$}
\State $w = \mathbf{w}[k_h,k_w]$
\State $f_h = \textbf{loadOffset}(k_h)$
\State $f_w = \textbf{loadOffset}(k_w)$
\For{$\widehat{o}_h = 0$, $\widehat{o}_h{+=}S$, while $\widehat{o}_h < T_{O_H}$}
\For{$\widehat{o}_w = 0$, $\widehat{o}_w{+=}S$, while $\widehat{o}_w < T_{O_W}$}
\State $o_h = \widehat{o}_h + f_h$
\State $o_w = \widehat{o}_w + f_w$
\State $i_h = ( o_h + P - k_h ) / S$
\State $i_w = ( o_w + P - k_w ) / S$
\State $\mathbf{y}[o_h,o_w] \leftarrow w \times \mathbf{x}[i_h,i_w]$
\EndFor
\EndFor
\EndFor
\EndFor
\State $\textbf{pushOutputBlock}(\mathbf{y})$
\EndFor
}
\end{algorithmic}
\caption{\protect\small{\textbf{Deconvolution Kernel.} Each kernel loads inputs, weights, and offsets into local memory to compute each output block. }}
 \label{alg:revd}
\end{algorithm}

\vspace{-0.25cm}
\noindent \textbf{(2) Dataflow Optimization.} Loop interchange is an algorithm-level optimization that can be applied to improve the sequential computation order of operations~\cite{ma2017optimizing}.
We reorder the loops of the deconvolution arithmetic in~\cite{zhang2017design} to sequentially traverse the weight space and maximize data reuse.
Increasing weight-level data reuse also increases the impact of zero-skipping - a conditional execution paradigm that eliminates redundant operations by only processing non-zero elements.

Additionally, we exploit the opportunities for data-level parallelism exposed by directly looping over the output space.
Unlike the standard deconvolution algorithm, which suffers from the overlapping sum problem, the output space of the reverse looping deconvolution can be tiled into smaller batches to execute concurrently on a parallelized hardware architecture.
When the size of the output feature space increases owing to the
upsampling nature of deconvolution operations, the
workloads and memory requirements remain constant,
simplifying hardware design requirements. \\

\vspace{-0.25cm}
\noindent \textbf{(3) Decoupling external memory accesses from compute operations.}
Reverse looping deconvolution arithmetic using~\cite{zhang2017design} produces a non-sequential external memory access pattern over the input space.
To mask any resulting overhead, we decouple all external memory accesses from compute operations to allow for the cascaded execution of these sub-tasks on a pipelined hardware architecture and restrict non-sequential memory access patterns to faster on-chip memory.
This is done by first computing the pixel addresses of an input block using Eq.~\ref{eq:input-pixel-calc-final}, then sequentially reading these addresses from external memory, and finally caching the data on-chip to be distributed.
To do this, we determine the tile size $T_{I_H}$ of the input block needed for each output block from the output tiling factor $T_{O_H}$ and the layer parameters using Eq.~\ref{eq:finding-tih-size}.
The resulting deconvolution kernel given by Algorithm~\ref{alg:revd} can then continuously compute $T_{O_H} \times T_{O_W}$ output blocks with a non-sequential access pattern over locally cached $T_{I_H} \times T_{I_W}$ input blocks using $K \times K$ weight blocks as the next
set of inputs are fetched from external memory using sequential reads.
\begin{equation}
\small{T_{I_H} = \max(i_h) - \min(i_h) = \ceil*{\frac{T_{O_H}}{S}} + \ceil*{\frac{K}{S}}}
\label{eq:finding-tih-size}
\end{equation}


\section{FPGA Hardware Architecture}
\label{sec:hardware-architecture}



To accelerate DCNN inference on an FPGA, we design a SIMD (Single Instruction Multiple Data) hardware architecture with replicable compute units (CUs) that exploits the opportunities for both spatial and temporal data-level parallelism that arise from the optimizations discussed in Section~\ref{sec:deconv-algorithm}.
As depicted in Figure~\ref{fig:hardware-architecture}, the dataflow of the deconvolution accelerator IP block is split into the three pipelined stages outlined below. \\

\vspace{-0.3cm}
\noindent \textbf{(1) Reading Inputs and Weights. }
The limited amount of on-chip memory is a bottleneck when accelerating large networks on a resource-limited FPGA.
As such, the input feature maps and network weights are stored in off-chip DDR memory and fetched using AXI interconnects.
As described in Section~\ref{sec:deconv-algorithm}, decoupling external memory accesses masks the communication overhead when executed in a pipelined architecture.
We separate input and weight external memory accesses into dedicated hardware blocks to concurrently read from DDR memory and stream to CUs through on-chip FIFOs.
This efficient memory hierarchy is realized by on-chip buffers using BRAMs to store tiled input and weight blocks to be processed by CUs.
\\

\vspace{-0.3cm}
\noindent \textbf{(2) Spatially Parallelized Compute Units. }
Looping over the output feature map enables partitioning deconvolution arithmetic into tiled batches that can execute concurrently across an array of CUs. 
The CUs follow a SIMD execution model, where each workload is dependent on blocks of inputs and weights that are sequentially streamed in through FIFOs and accumulated.
The CUs each perform the deconvolution arithmetic outlined in Algorithm~\ref{alg:revd} using on-chip DSP units and the resulting $T_{O_H} \times T_{O_W}$ output block is streamed out to be written to off-chip memory.
To maximize the occupancy of these CUs, we explore the design space as outlined in Section~\ref{sec:design-space-exploration} to optimize the output tiling factor.
\\

\vspace{-0.3cm}
\noindent \textbf{(3) Writing Output Pixels. } Traversing the output space and avoiding the overlapping sum problem allows for a one-shot write to external memory for each output block computed by a CU.
We dedicate a hardware block to stream the outputs from each element in the CU array to be written to external DDR memory.
This minimizes communication overhead with DDR and on-chip BRAM memory requirements. \\

\begin{figure}
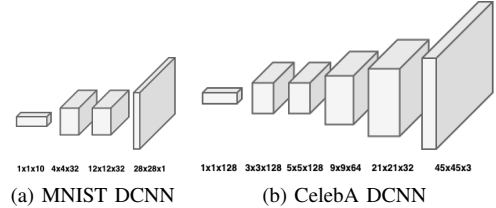

    \centering
    \subfloat[MNIST DCNN]{
    \includegraphics[width=0.25\linewidth]{FIGS/MNIST-GAN.png}
    }
    \subfloat[CelebA DCNN]{
    \includegraphics[width=0.45\linewidth]{FIGS/CelebA-GAN.png}
    }
    \caption{\small{\textbf{DCNN Architectures.} We consider the network architectures shown above for inference acceleration on low-power hardware.}}
    \label{fig:architecture}
\end{figure}

\vspace{-0.3cm}
\section{Experimental Results}
\label{sec:roofline-model}
We implement our architecture on a Xilinx PYNQ-Z2 board at 32-bit fixed point precision using the Vivado Design Suite.
With the available hardware resources, we synthesize the design with 16 CUs at 125MHz in Vivado HLS using HLSLIB~\cite{hlslib} and benchmark performance on the two DCNNs depicted in Figure~\ref{fig:architecture}. Each DCNN is trained on the MNIST and CelebA datasets using the WGAN-GP~\cite{gulrajani2017improved} framework.

\subsection{Design Space Exploration}
\label{sec:design-space-exploration}

\begin{figure}[t]
    \centering
    \subfloat[MNIST DCNN]{
    \includegraphics[width=0.4\linewidth]{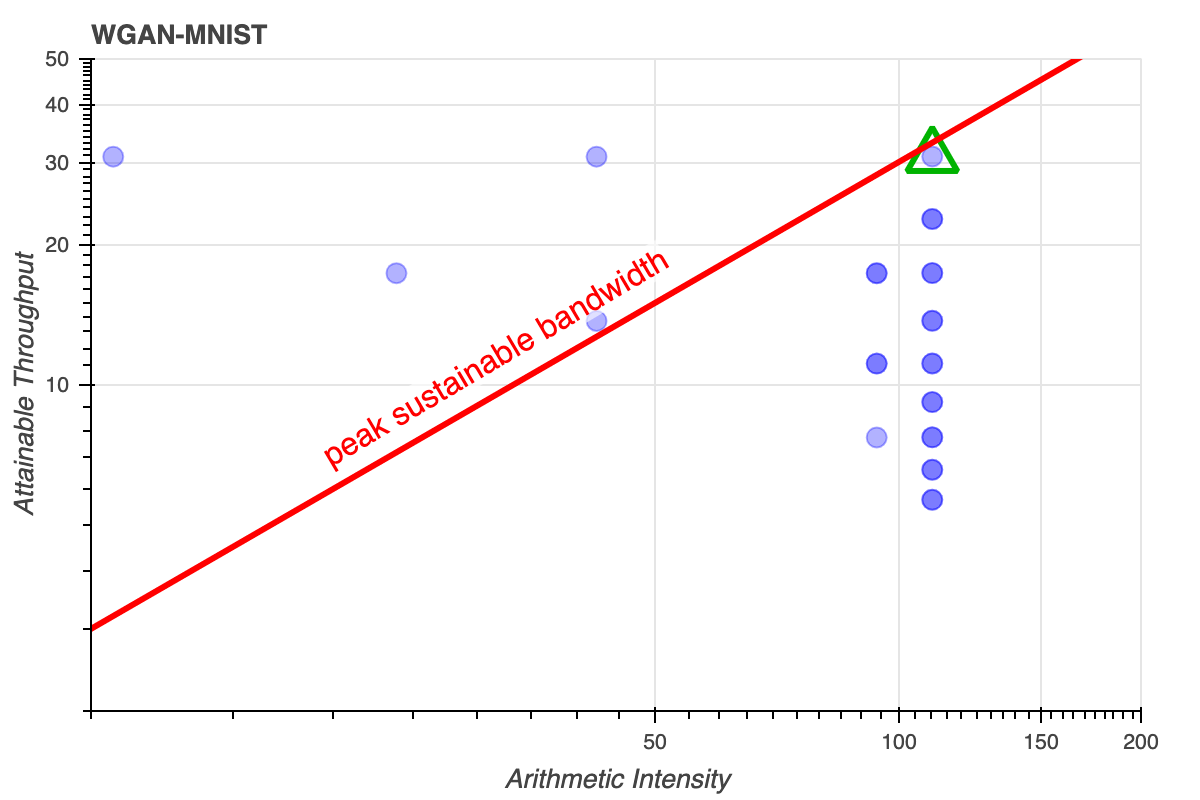}
    }
    \subfloat[CelebA DCNN]{
    \includegraphics[width=0.4\linewidth]{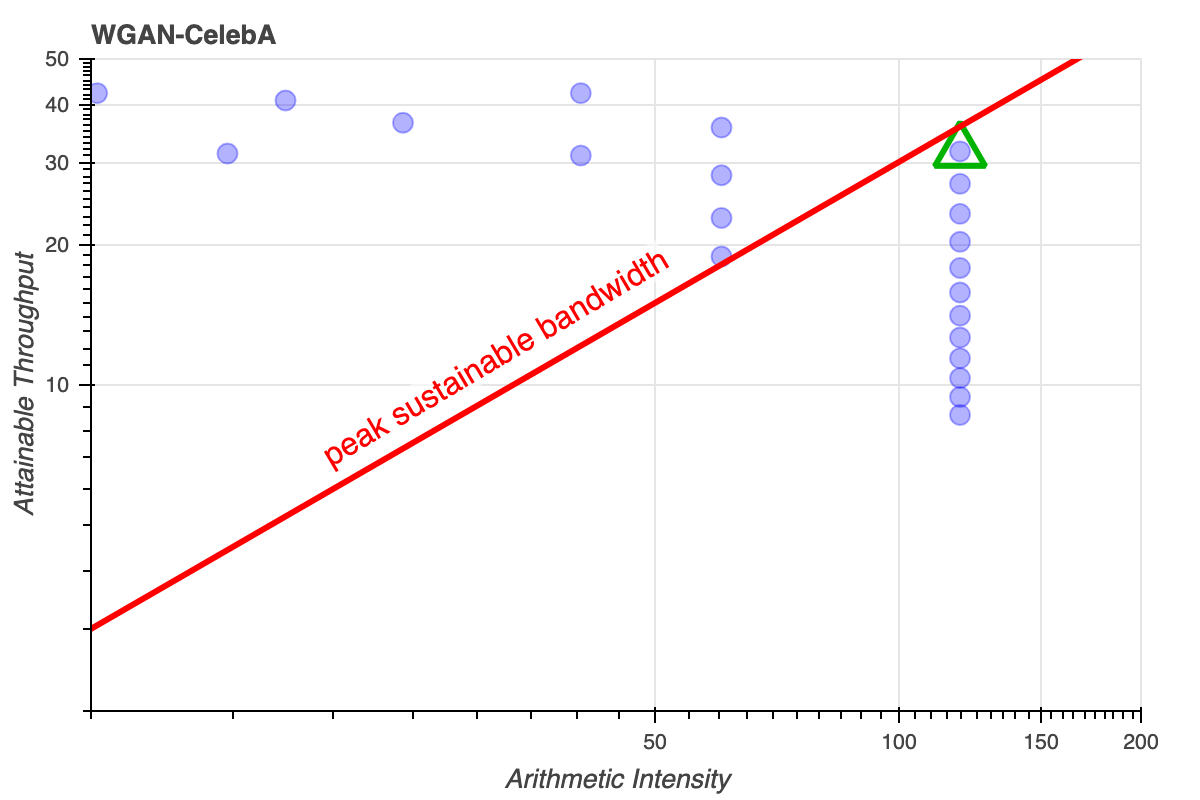}
    }

    \caption{\small{\textbf{Design Space Exploration.} The optimal tiling factor $T_{O_H}$ maximizes attainable throughput while satisfying the peak sustainable bandwidth constraint as measured by the STREAM benchmark~\cite{mccalpin1995stream}.}}
    \label{fig:design-exploration}
\end{figure}

In this work, we explore square tiling factors over the output space such that $T_{O_H} = T_{O_W}$ and use the design space exploration methodology proposed by Zhang \textit{et al.}~\cite{zhang2015optimizing} to optimize $T_{O_H}$. Because our accelerator multiplexes through the DCNN layers, we optimize $T_{O_H}$ globally across all layers for each network architecture as a unified hardware design parameter as in~\cite{zhang2015optimizing}. Fig.~\ref{fig:design-exploration} depicts all legal solutions for both the MNIST and CelebA DCNNs. 
Any solution to the left of the peak sustainable bandwidth slope requires a higher bandwidth than the FPGA can sustain~\cite{zhang2015optimizing}.
The optimal $T_{O_H}$ (indicated in green) maximizes attainable throughput while satisfying the hardware constraints. Table~\ref{tbl:hardware-utilization} provides the values used in this work and the resulting FPGA resource utilization.
Note that the Xilinx PYNQ-Z2 board is extremely resource-constained, using only 9\% of the DSP blocks used in~\cite{yazdanbakhsh2018flexigan} and 5\% of that used in~\cite{wang2019towards} and~\cite{chang2020towards}.

\begin{table}[t]
\begin{center}
\fontsize{6}{8}\selectfont
\begin{tabular}{lccccc}
\hline
\rowcolor[HTML]{EFEFEF} 
\multicolumn{1}{c}{} & $T_{O_H}$ & DSP48s & BRAMs & Flip-Flops & LUTs \\ \hline
\multicolumn{1}{|l|}{MNIST} & \multicolumn{1}{c|}{12} & \multicolumn{1}{c|}{134} & \multicolumn{1}{c|}{50} & \multicolumn{1}{c|}{43218} & \multicolumn{1}{c|}{36469} \\ \hline
\multicolumn{1}{|l|}{CelebA} & \multicolumn{1}{c|}{24} & \multicolumn{1}{c|}{134} & \multicolumn{1}{c|}{74} & \multicolumn{1}{c|}{48938} & \multicolumn{1}{c|}{40923} \\ \hline
\end{tabular}
\caption{\small{\textbf{Xilinx PYNQ-Z2 Resource Utilization}}}
\label{tbl:hardware-utilization}
\end{center}
\end{table}

\vspace{-0.2cm}
\subsection{Performance-per-Watt Comparison with Edge GPU}
\label{sec:perf-analysis}

\begin{table}[b!]

\begin{center}
\fontsize{6}{8}\selectfont

\begin{tabular}{lcccl}
\hline
\rowcolor[HTML]{EFEFEF} 
\multicolumn{1}{c}{\textbf{MNIST}} & L1 & L2 & L3 & \multicolumn{1}{c}{Total} \\ \hline
\multicolumn{1}{|l|}{FPGA} & \multicolumn{1}{c|}{\textbf{2.4 (0.02)}} & \multicolumn{1}{c|}{\textbf{3.0 (0.01)}} & \multicolumn{1}{c||}{\textbf{2.8 (0.01)}} & \multicolumn{1}{l|}{\textbf{2.9 (0.01)}} \\ \hline
\multicolumn{1}{|l|}{GPU} & \multicolumn{1}{c|}{1.3 (0.17)} & \multicolumn{1}{c|}{2.7 (0.42)} & \multicolumn{1}{c||}{1.8 (0.25)} & \multicolumn{1}{l|}{2.1 (0.18)} \\ \hline
\end{tabular}

\bigskip
\begin{tabular}{lcccccl}
\hline
\rowcolor[HTML]{EFEFEF} 
\multicolumn{1}{c}{\textbf{CelebA}} & L1 & L2 & L3 & L4 & L5 & \multicolumn{1}{c}{Total} \\ \hline
\multicolumn{1}{|l|}{FPGA} & \multicolumn{1}{c|}{\textbf{4.0 (0.00)}} & \multicolumn{1}{c|}{4.0 (0.00)} & \multicolumn{1}{c|}{\textbf{4.0 (0.00)}} & \multicolumn{1}{c|}{2.3 (0.00)} & \multicolumn{1}{c||}{1.2 (0.01)} & \multicolumn{1}{l|}{\textbf{3.9 (0.00)}} \\ \hline
\multicolumn{1}{|l|}{GPU} & \multicolumn{1}{c|}{3.2 (0.66)} & \multicolumn{1}{c|}{\textbf{4.4 (0.81)}} & \multicolumn{1}{c|}{3.9 (0.66)} & \multicolumn{1}{c|}{\textbf{4.4 (0.69)}} & \multicolumn{1}{c||}{\textbf{2.2 (0.40)}} & \multicolumn{1}{l|}{3.6 (0.31)} \\ \hline
\end{tabular}

\caption{\small{\textbf{DCNN Comparison (GOps/second/Watt).} We measure the mean and standard dev. (in parenthesis) of the throughput to power ratio of each layer in each DCNN on each processor over 50 runs.}}
\label{tbl:fpga-gpu-comp}
\end{center}
\end{table}

GPUs are power-hungry processors heavily optimized for large batch processing of on-chip memory~\cite{harris2005mapping}.
\ianNotes{Unlike the FPGA, which has been shown to provide workload-insensitive throughput with better power-efficiency, the time-varying optimizations leveraged by modern GPUs give rise to a non-deterministic execution model that can rarely provide the consistent performance that is required by edge computing applications~\cite{jouppi2017datacenter,biookaghazadeh2018fpgas}.} 
Additionally, modern GPUs use hardware throttling (ie. reducing clock frequency) to lower power and cool the chip when it gets hot, further increasing run-to-run variation~\cite{jetson2020guide}.
This makes FPGAs the more suitable choice for edge computing applications when consistent throughput and power efficiency are key requirements~\cite{biookaghazadeh2018fpgas}.


\ianNotes{In our experiments, we compare the throughput to power ratio of our Xilinx PYNQ-Z2 FPGA design against the heavily optimized NVIDIA Jetson TX1 edge computing GPU.}
As in~\cite{nurvitadhi2017can}, we evaluate the GPU with Torch using nvprof to collect performance and power numbers for each layer in each DCNN.
We measure FPGA power using a USB Power Meter Voltage Detector and collect performance numbers using hardware counters.
We compute total network throughput as the sum of the arithmetic operations of all layers divided by the sum of the execution time of all layers.
Our results provided in Table~\ref{tbl:fpga-gpu-comp} show that our design yields a higher total network throughput to power ratio with lower run-to-run variation when compared to the GPU for both DCNNs.
As noted in~\cite{zhang2015optimizing}, unified design parameters such as $T_{O_H}$ simplify implementation cost but may be sub-optimal for some layers.
We observe this behavior for the CelebA DCNN as shown in Table~\ref{tbl:fpga-gpu-comp}.
In future work, we will investigate dynamically reconfiguring tiling factors to optimize dataflow per layer.

\begin{figure}[t]
    \centering
    \subfloat[Hardware Performance]{
    \includegraphics[width=0.35\linewidth]{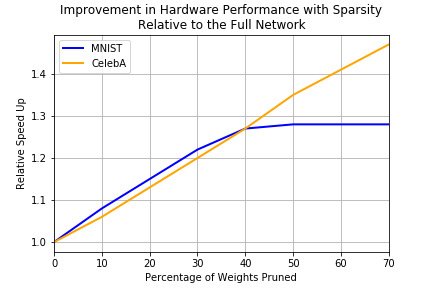}
    }
    \subfloat[Generative Quality]{
    \includegraphics[width=0.35\linewidth]{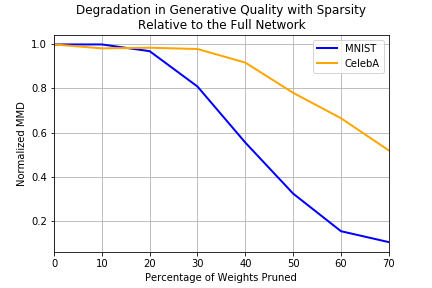}
    } \\
    \subfloat[Balancing the Trade-Off]{
    \includegraphics[width=0.48\linewidth]{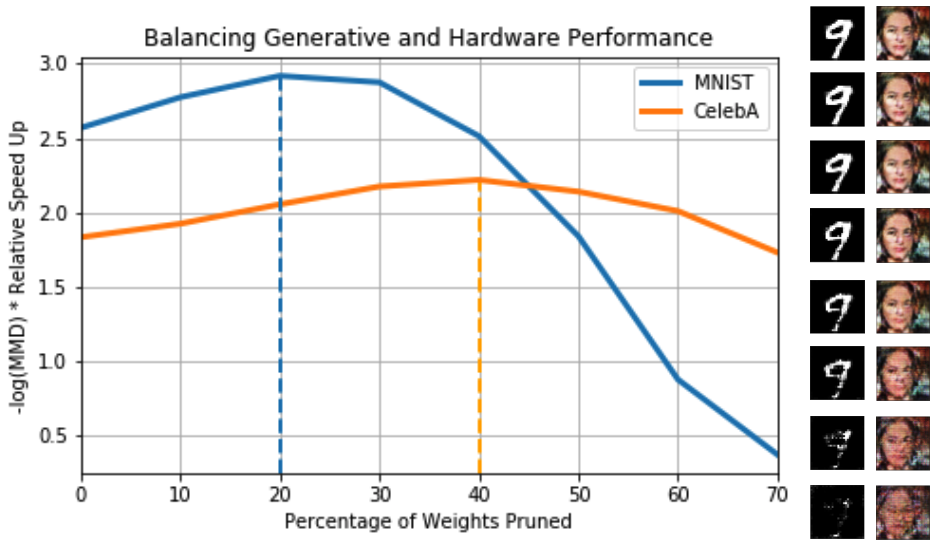}
    }
    \caption{\small{\textbf{FPGA Sparsity Analysis.} As described in Section~\ref{sec:sparsity}, while unstructured sparsity leads to FPGA speed-ups when using conditional execution paradigms like zero-skipping, removing learned parameters from a network invariably leads to degradation in generative quality. We propose a design metric to balance this trade-off.}}
    \label{fig:sparsity-analysis}
\end{figure}

\vspace{-0.3cm}
\subsection{Sparsity Experiments}
\label{sec:sparsity}
Weight pruning is a widely studied technique used to reduce network power consumption and memory footprint on mobile and edge computing platforms~\cite{han2015deep}.
It's difficult for GPUs to effectively accelerate this form of unstructured sparsity as they are highly sensitive to conditional execution paradigms such as zero-skipping~\cite{biookaghazadeh2018fpgas,nurvitadhi2017can}.
Alternatively, FPGA performance is stable under such paradigms and can yield significant speed-ups when only executing non-zero valued computations~\cite{biookaghazadeh2018fpgas,chang2019sdcnn}.
Previous works optimizing DCNN dataflow for unstructured sparsity fail to account for this degradation~\cite{chang2019sdcnn}.

In our experiments, we systematically prune DCNN weights by their magnitude as done in~\cite{han2015deep}.
To visualize both hardware performance and generative quality, we analyze the rates of change of both system latency and Maximum Mean Discrepancy (MMD) distance, respectively. MMD distance is used to compute the dissimilarity between model-parameterized distribution $P_\theta$ and ground truth distribution $P_g$ and, in practice, is empirically estimated by drawing independent samples $\{ \bm{x_1}, \cdots, \bm{x_n} \} \sim \mu$ and $\{ \bm{y_1}, \cdots, \bm{y_n} \} \sim \nu$ from distributions $P_\theta$ and $P_g$, respectively, where kernel $k$ maps to a reproducing kernel Hilbert space~\cite{borji2019pros,gretton2012kernel}.
It the follows that the MMD distance given by the equation below is zero if and only if the distributions are identical.
Here, we explore the use of MMD with the standard Gaussian kernel $k(\bm{x}, \bm{x}') = \exp( \Vert \bm{x} - \bm{x}' \Vert^2 )$ using the Euclidean distance, selecting the median euclidean distance between ground truth samples as the bandwidth~\cite{gretton2012kernel}.

\vspace{-0.3cm}
{\small
$$
    \text{MMD}_k(\mu,\nu) = \mathbb{E}_{\mu, \mu} [ k(X,X') ] + \mathbb{E}_{ \nu, \nu } [ k(Y,Y') ] -  2 \mathbb{E}_{ \mu, \nu} [ k(X,Y) ] 
$$
}

\vspace{-0.3cm}
Pruning more weights yields higher speed-ups when skipping computations with zero-valued weights, as shown in Fig~\ref{fig:sparsity-analysis}-a.
However, as shown in Fig~\ref{fig:sparsity-analysis}-b, the generative quality decreases with added sparsity.
To balance the trade-off between hardware performance and generative quality, we propose an optimization metric given by Eq.~\ref{eq:mmd-metric}.
Here, $t_0$ and $d_0$ denote the execution time and MMD distance with respect $P_g$ using the full weight matrix $\theta_0$ while $t_p$ and $d_p$ denote that of the sparse matrix $\theta_p$ where $d_i$ is given by $\text{MMD}( P_g, P_{\theta_i} )$.
Multiplying the rate of change of system latency and MMD distance leads to a concave optimization curve with a peak representing the sparsity level that balances image quality with execution time.
\vspace{-0.2cm}
\begin{equation}
\small{
    \frac{d_0}{d_p} \times \frac{t_0}{t_p}
}
\label{eq:mmd-metric}
\end{equation}

\vspace{-0.1cm}



\section{Conclusions and Future Work}
In this paper, we adapt the deconvolution algorithm first proposed in~\cite{zhang2017design} to a parallelized execution model by reducing resource utilization, improving dataflow, and exploiting memory hierarchy.
We design a spatio-temporally parallelized hardware architecture to accelerate this algorithm for DCNN inference on a Xilinx PYNQ-Z2 FPGA.
For edge computing applications when consistent throughput and power efficiency are key requirements, we show that this resource-limited FPGA achieves a higher throughput to power ratio with lower run-to-run variation than the NVIDIA Jetson TX1 edge computing GPU.
To balance the trade-off between generative quality and hardware performance, we propose a MMD-based optimization metric when exploring unstructured sparsity.
In future work, we will adapt this architecture to other GANs and investigate the effect of bitwidth reduction on hardware performance and generative quality.

\vspace{-0.2cm}
\bigskip
\centerline{A\footnotesize{CKNOWLEDGEMENTS}}

\medskip
This work was supported in part by NSF awards CNS-1730158, ACI-1540112,  ACI-1541349, OAC-1826967, the University of California Office of the President, and the California Institute for Telecommunications and Information Technology's Qualcomm Institute (Calit2-QI). We would also like to thank Parimal Patel and Stephen Neuendorffer at Xilinx and Byungheon Jeon at UC San Diego.

\bibliography{citations.bib}

\end{document}